# Accelerator magnet development based on COMB technology with STAR® wires

V V Kashikhin[1], S Cohan[1], V Lombardo[1], D Turrioni[1], N Mai[2], A K Chavda[2], U Sambangi[2], S Korupolu[2], J Peram[2], A Anil[2], C Goel[2], J Sai Sandra[2], V Yerraguravagari[2], R Schmidt[2], V Selvamanickam[2], G Majkic[2], E Galstyan[3] and K Selvamanickam[3]

[1]Fermi National Accelerator Laboratory
[2]Department of Mechanical Engineering, Advanced Manufacturing Institute, University of Houston
[3]AMPeers LLC

Email: vadim@fnal.gov

**Abstract**. This paper reports progress in the development of COMB magnet technology with STAR® wires. A two-layer dipole magnet with 60 mm clear bore has been recently fabricated and tested in liquid nitrogen. The purpose of the test was to determine what kind of critical current degradation occurs in the process of winding the STAR® wire into the COMB structure.

## 1. Introduction

Fermilab is working on superconducting accelerator magnet R&D under the framework of the U.S. Magnet Development Program (USMDP) [1]. An integral part of that program is High Temperature Superconducting (HTS) accelerator magnet development to demonstrate self-fields of 5 T or greater compatible with operation in hybrid LTS/HTS configuration to generate fields beyond 16 T for future High Energy Physics (HEP) experiments. The ever-increasing requirements on the magnetic field strength from the physics community lead to exorbitant levels of mechanical stresses in the coils, placing limits on the use of RE-Ba$_2$Cu$_3$O$_7$−δ (REBCO) coated conductors – one of the most promising HTS materials for high field accelerator magnets [2].

To address this issue, Fermilab is developing the Conductor on Molded Barrel (COMB) magnet technology, which offers an elegant solution for fabrication of dipole, quadrupole and higher-pole coils with round HTS conductors, offering stress management and precise turn positioning to control the magnetic field quality [3]. The COMB magnet technology couples well with Symmetric Tape Round (STAR®) wires produced by AMPeers LLC, which are state-of-the-art multi-tape REBCO conductors. Due to the proprietary tape architecture placing the superconducting layer near the neutral plane of the tape, they offer unrivaled bending performance suitable for high field magnets with the bore size in 50-60 mm range, which will be needed for future HEP experiments [4], [5].

One of the most pressing questions for HTS application in accelerator magnets is what kind of the critical current ($I_c$) degradation can occur in the process of winding the conductor into the support structure with a relevant curvature. Short STAR® wire measurements performed in the past [6], [7] indicated that the critical current retention was as high as 97% for a 30 mm bending diameter. However, general experience with winding other REBCO conductors suggests that the actual $I_c$ retention due to winding the conductor into the structure may be considerably lower than during short (i.e. hairpin) sample tests due to various contributing factors. Therefore, a wire winding and testing experiment, described in this paper, was needed to draw a definitive conclusion on whether the STAR® wires can withstand the coil winding process and retain their performance.

## 2. STAR® wire design, fabrication and testing

STAR® wire fabrication for this project followed the process sequence shown in figure 1. All fabricated STAR® wires used REBCO tapes made by Advanced Metal Organic Chemical Vapor Deposition (MOCVD) [8] as the starting material. The tapes consisted of 4-µm-thick (Gd,Y)-Ba-Cu-O films with 5% Zr addition. The critical current of the long tapes was measured by reel-to-reel Scanning Hall Probe Microscopy (SHPM). The magnetic field dependence of critical current of the two 50-m-long A-MOCVD tapes used in this work was measured using a Physical Property Measurement System (PPMS) at 77 K and 4.2 K and the results are displayed in figure 2.

This data was used later to determine the expected critical current STAR® wires made with the Advanced MOCVD REBCO tape strands, in a flat form, after winding on a test holder and after winding into a COMB magnet. At 4.2 K, 13 T, the lift factor in critical current of Tapes 1 and 2 was similar at 4.06 and 4.43 respectively. The critical current of Tape 1 and Tape 2 at 4.2 K, 13 T are then 1,266 A/4mm and 1,245 A/4mm respectively. These values are 2.5x the critical current of typical commercial REBCO tapes at 4.2 K, 13 T.

12-mm-wide REBCO tapes with a total thickness of ~25 µm were laser slit to 2- and 2.6-mm widths. Generally, four narrow tape strands (one 2 mm and three 2.6 mm) were obtained from a single 12 mm wide tape. A thin layer of silver was sputter deposited to seal the slit edges of the strands. After oxygenation, copper stabilizer was electroplated primarily on the REBCO film side so as to position the film near the neutral plane.

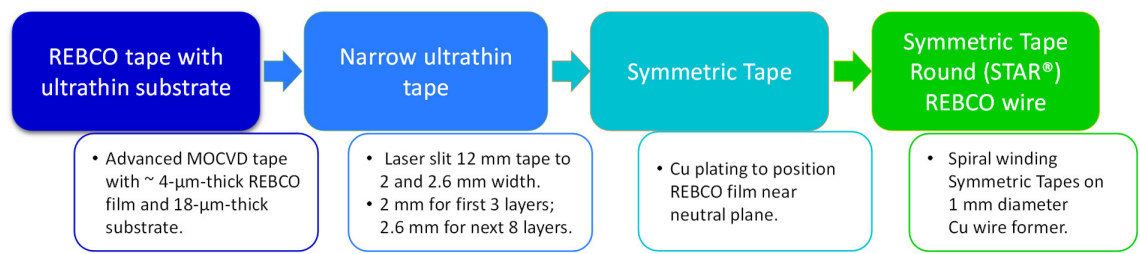

**Figure 1**. STAR® wire fabrication process.

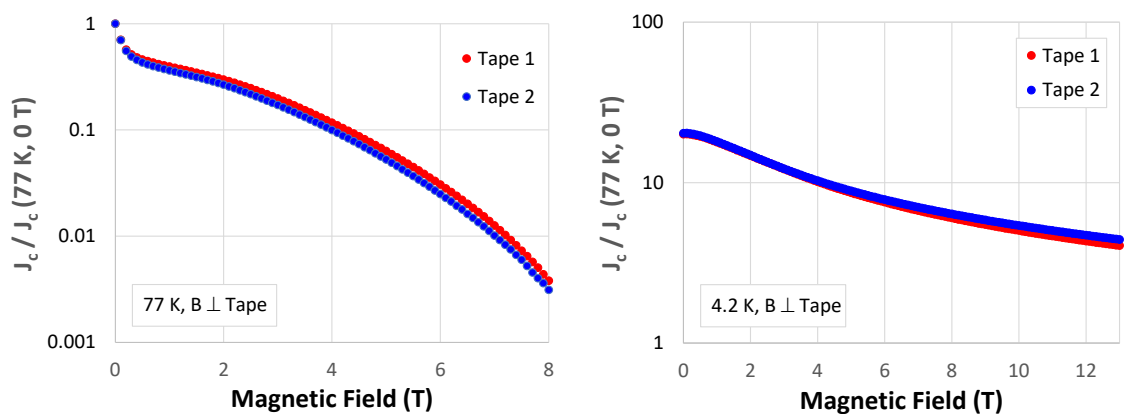

**Figure 2.** Normalized magnetic field dependence of $I_c$ at 77 K (left) and lift factor in $I_c$ at 4.2 K of two 50-m-long, 12-mm-wide REBCO tapes made with 4-µm-thick films by Advanced MOCVD for STAR® wire fabrication (right).

The optimum copper thickness was confirmed by spiral winding the plated tape on a 1 mm diameter former and measuring the $I_c$ retention. On confirming the optimum copper thickness, the long, slit tapes were electroplated with copper. The thickness of tape strands after copper plating was 52 – 54 µm. The $I_c$ of the narrow tape strands was tested again by SHPM at 77 K, 0 T. For each of the two 5-m-long STAR® wires, three 2-mm-wide tape strands and eight 2.6-mm-wide tape strands were selected based on the SHPM data. Each of the 11 strands was 8 meters in length. A 1.02-mm copper former was used to wind three 2-mm-wide and eight 2.6-mm-wide tape strands. The STAR® wires had designations STAR202303131 and STAR202303151, which will be further referred to as STAR® wires 131 and 151. The diameters of STAR® wires 131 and 151 were 2.4 mm and 2.34 mm respectively.

The wire ends were terminated with ¼ inch outer diameter and 200-mm long tubes from C10100 oxygen-free copper. The tubes were perforated on one side and filled with liquid indium in horizontal position, soldering the STAR® wire inside. There were two voltage taps installed at each wire end. One voltage tap was attached to the outermost layer of the tapes, 1 cm inside the terminal, and the other voltage tap was attached to the prior layer, 2 cm inside of the terminal. The distance between the voltage taps was about 4.7 m.

The STAR® wires were tested stand-alone in liquid nitrogen prior to the magnet fabrication in order to determine $I_c$ of undisturbed wires. A sample holder, shown in figure 3, was fabricated for that purpose. It accommodated 5.5 turns of the STAR® wire wound into a groove with 250 mm wire supporting diameter. The magnetic field on the conductor is shown in figure 3 as well. The peak magnetic field, located at the end turns was fairly uniformly distributed among them. It means that about 30% of the conductor was exposed to the peak field level with the transfer function of 0.192 T/kA, which was slightly larger than a theoretical transfer function of the straight wire of 0.143 T/kA.

Figure 4 shows typical voltage-current characteristics of two wires tested in liquid nitrogen. The wires exhibited resistances of 100-200 nΩ starting from low currents, which were removed for the purposes of determining the $I_c$ and n-value. The performance of two wires was quite similar with the minimum critical current in a 595-606 A range using the electrical field criterion of 0.4 µV/cm. The n-value was between 8-9 for both wires.

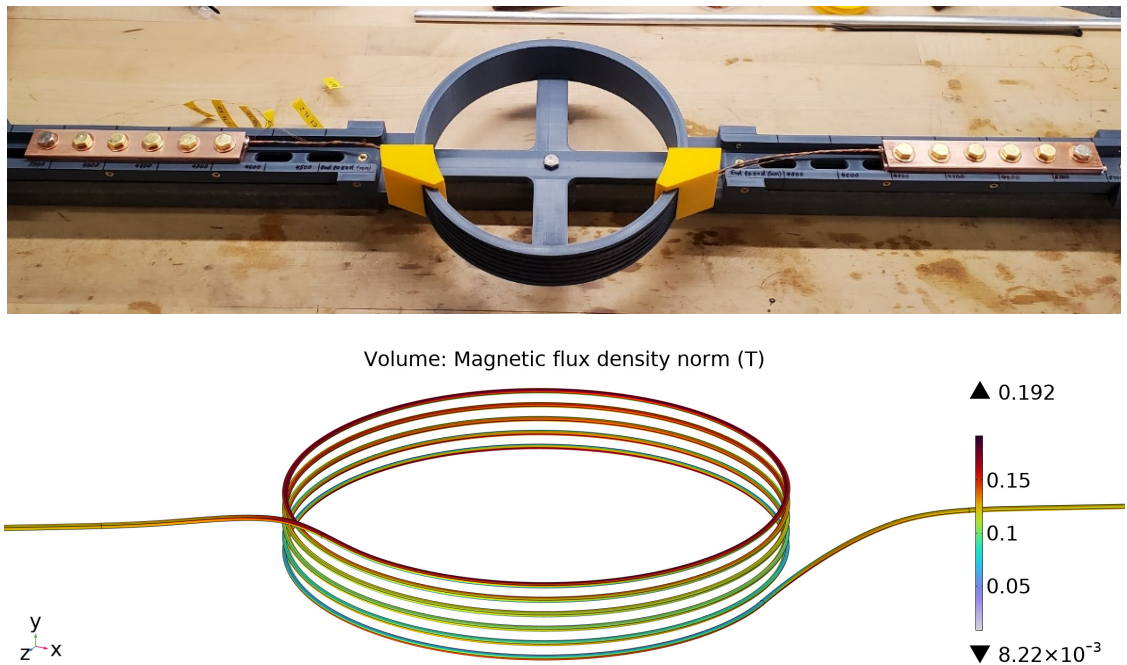

**Figure 3**. A 5-m long STAR® wire installed into the sample holder (top) and the magnetic field on the wire at 1 kA current (bottom).

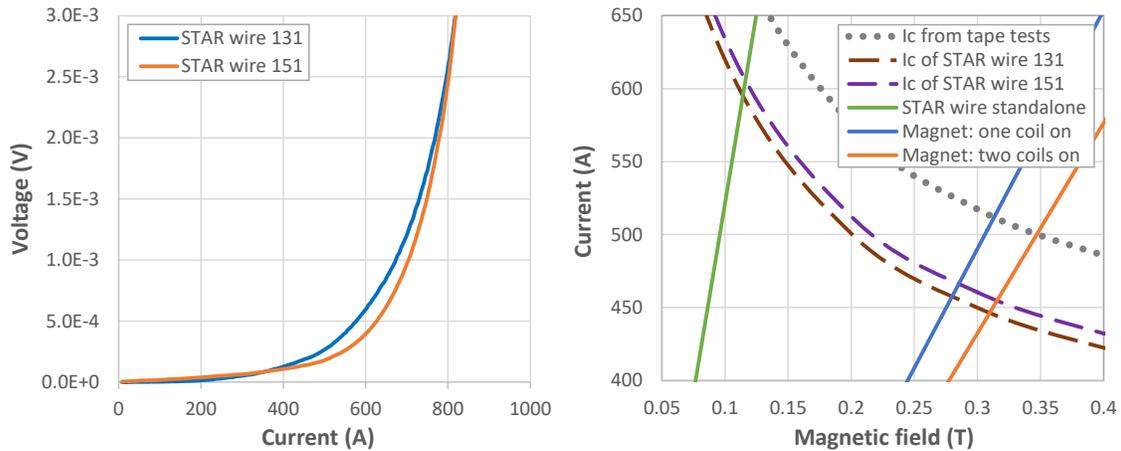

**Figure 4**. Typical voltages of STAR® wires tested standalone (left) and the expected $I_c$ (right).

The load line corresponding to the peak field in the sample holder is shown in figure 4, along with the expected critical current dependence vs. the magnetic field. The dotted curve is the sum of measured $I_c$ of all the tapes in the wire, measured during tape fabrication and shown in figure 2. The dashed curves are the above wire $I_c$ multiplied by scale factors of 0.87-0.89 to match the measured minimal critical currents at the intersection with the sample holder load line.

## 3. COMB-STAR-1 magnet design, fabrication and testing

### 3.1. Magnet design
The magnet, named COMB-STAR-1, had two half-coils holding two layers of cable each with no internal splice. The total length of the cable per half-coil was 4.76 m including the leads. The terminal tube on one end of each wire used for the stand-alone wire testing was cut off to facilitate the coil winding and allow proper trimming of the conductor length and re-installed after winding. Three copper adapters connected the half-coils together and to the power supply. The magnet had 60 mm clear bore and an iron yoke made of low carbon steel. Figure 5 shows the magnet solid model.

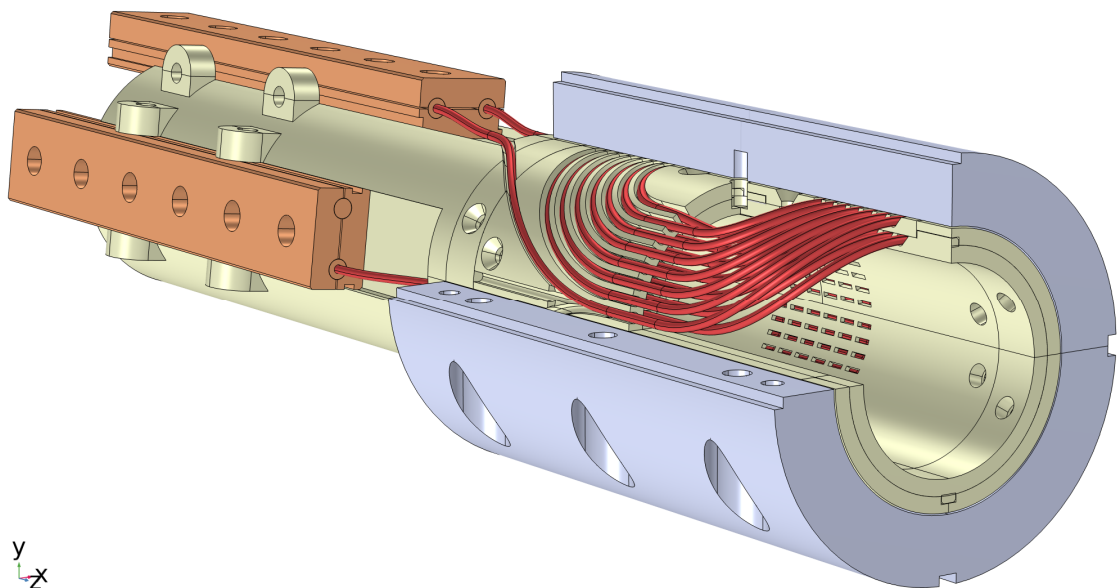

**Figure 5**. COMB-STAR-1 magnet solid model (a part of the structure is removed for clarity).

It was decided not to insulate the STAR® wires, to avoid locking in the tapes and to allow them to re-distribute freely during the coil winding, and to pair it with an electrically insulating support structure. ULTEM™ 1010 - a high strength thermoplastic frequently used in cryogenic applications was selected for the structure. The structure was designed with an oversized channel compared to the STAR® wire diameter to allow for thermal contraction during cool-down without straining the wire, and was fabricated by Fused Deposition Modelling (3D-printing) with a nearly 100% fill factor. Although the scope of this project was limited to testing the magnet in liquid nitrogen, the support structure was designed for considerably larger Lorentz forces in liquid helium as this test is foreseen in future.

Figure 6 shows the cross-section of one coil quadrant with the magnetic field distribution and the equivalent (von Mises) stress in the conductor and the support structure for the maximum current of 8.8 kA, corresponding to the bore field of 3.9 T, expected in superfluid helium at 1.9 K based on the wire $I_c$ measurements. Since the current and the magnetic field in liquid nitrogen is a factor of ~20 lower, the magnet had a smaller iron yoke exclusively for testing at 77 K and will later be outfitted with a larger iron yoke for testing in liquid helium.

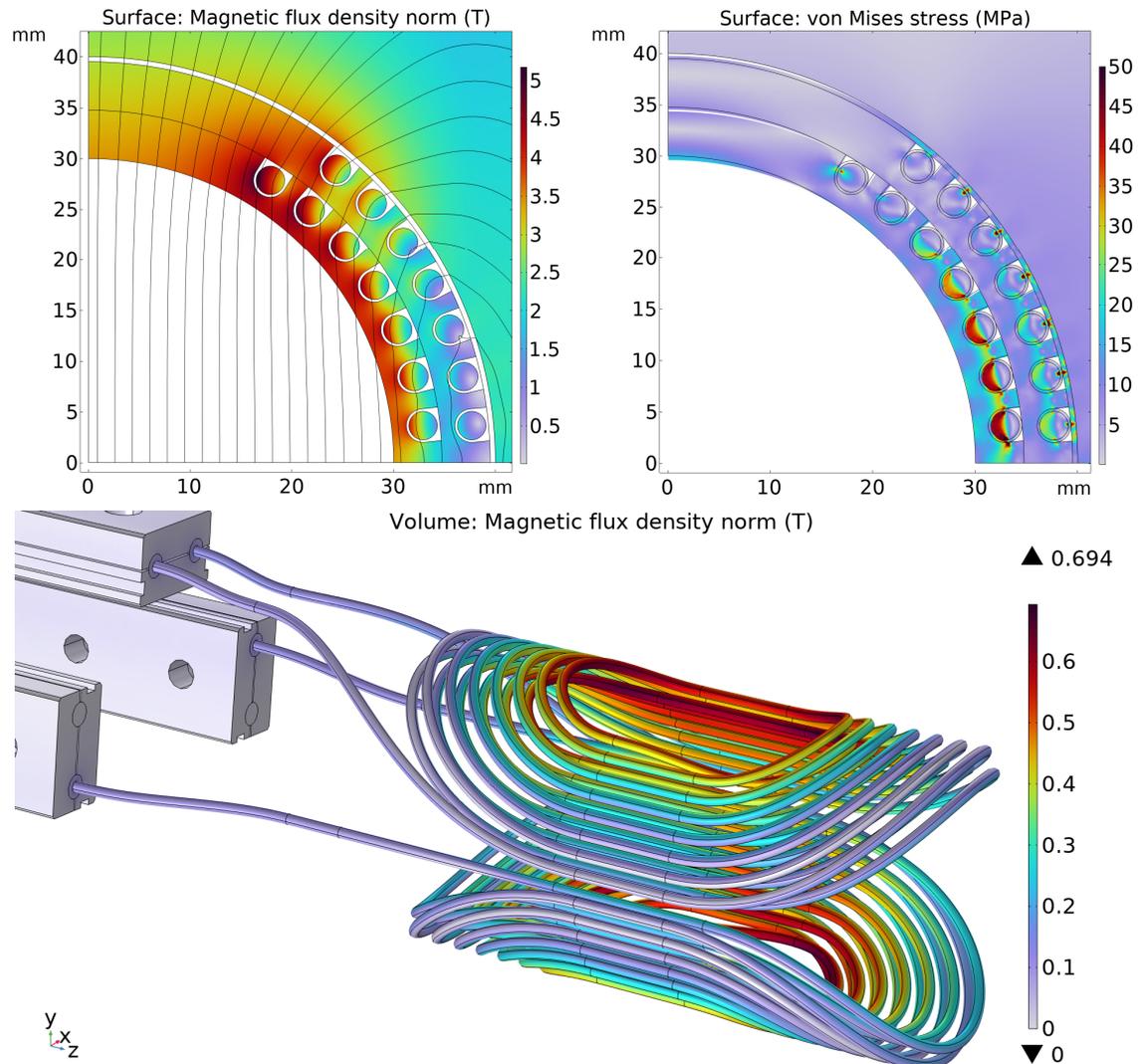

**Figure 6**. 2D magnetic field distribution (top-left) and the equivalent stress on the conductor and the support structure at 8.8 kA (top-right). 3D coil and lead geometry with magnetic field on the conductor at 1 kA (bottom); the iron yoke was present during calculations but not shown in the picture.

The peak equivalent stress in the structure calculated at 8.8 kA was around 50 MPa, which was significantly lower than the reported yield strength in compression of 245 MPa for the selected material. The peak stress in the conductor was slightly lower at 46 MPa for the midplane turn. No strain-related degradation of the conductor performance is expected at this stress level. The maximum displacement of the pole turn due to the structure deformation under the Lorentz Forces is 0.2 mm.

A complete 3D magnetic model was created as knowing the peak magnetic field on the conductor was necessary for the purposes of this project. Figure 6 shows the magnetic field on the conductor with the peak field at the pole turn of the inner layer. Since the iron yoke is not saturated during testing at 77 K, the peak field on the conductor can be conveniently characterized by a constant transfer function of 0.694 T/kA up to about 1 kA current. Intersection of the magnet load line with the measured wire characteristic in figure 2, gives the $I_c$ of about 450 A during the magnet test in liquid nitrogen. A more detailed analysis and comparison between the coils is presented in the magnet testing section.

### 3.2. Magnet fabrication

The magnet fabrication started from manually winding the voltage taps into the support structure of the inner layer of half-coil 1. It was then followed by manually winding the STAR® wire 131 into the same channel, as shown in figure 7, securing the second layer structure on top of the inner layer and proceeding with the voltage taps and conductor winding procedure for that layer. The most critical operation was winding the conductor around the poles of the inner layer with a diameter of ~32 mm – the narrowest bends in the magnet; it went without unexpected tape deformations or other issues.

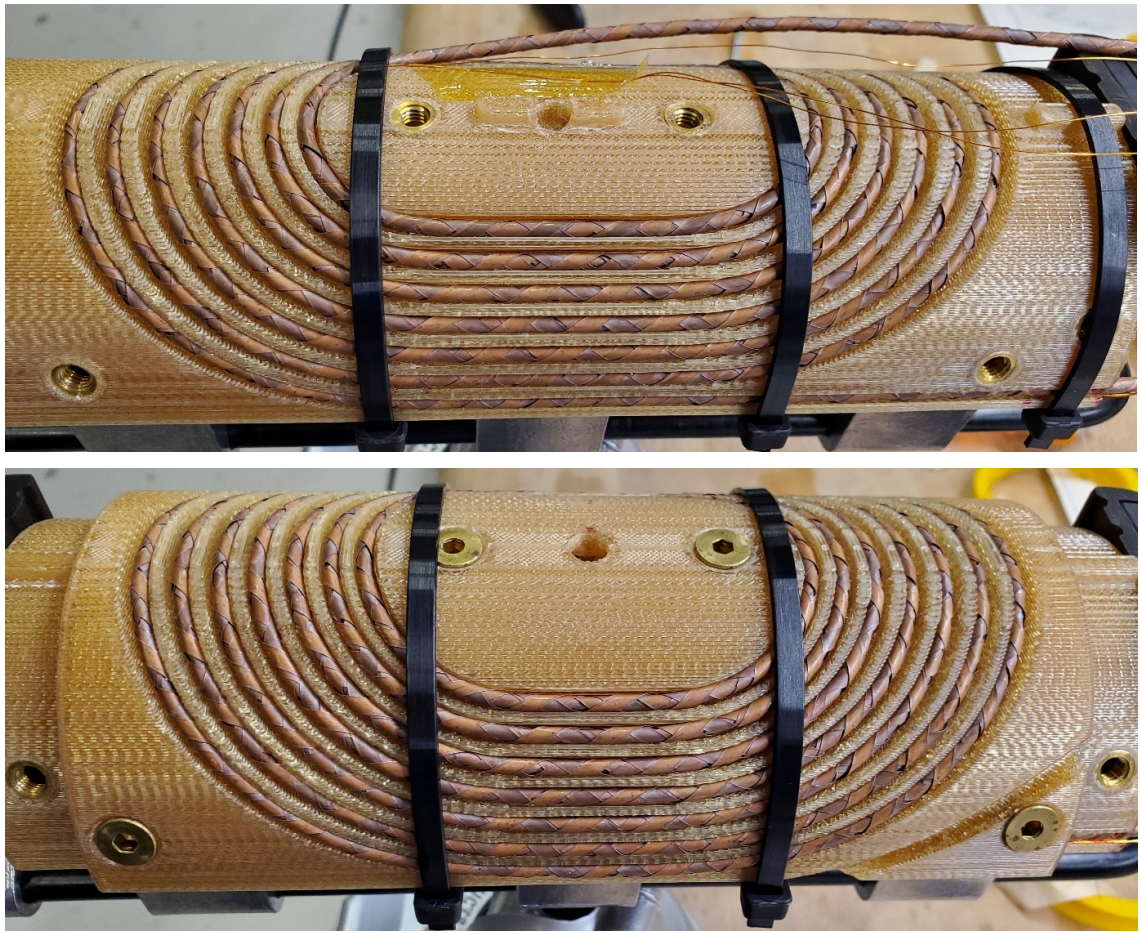

**Figure 7**. Inner (top) and outer (bottom) layers of half-coil 1 as wound.

The same steps were repeated for the half-coil 2 wound from the STAR® wire 151. Then the two half-coils were assembled with the lead support structure on one end and the retaining ring on the other. The ground insulation consisting of ~0.5 mm thick polyester film was installed around the coils as shown in figure 8. The tubes terminating the leads were secured inside of the copper adapters installed in the lead support. Then the coils were assembled with two halves of the iron yoke, which had alignment features between themselves as well as the coils.

The last step was to install a Hall probe array into the magnet bore. It consisted of three probes positioned to measure the dipole field component on the magnet axis with one probe placed at the magnet centre and the other two +/- 50 mm apart. Figure 8 shows the magnet assembled and ready for testing.

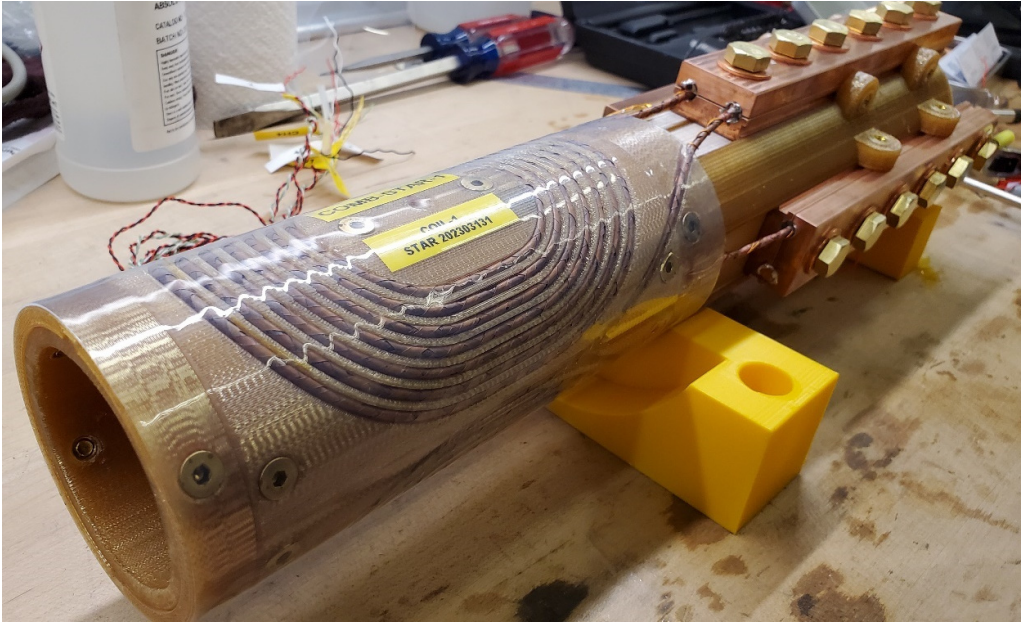

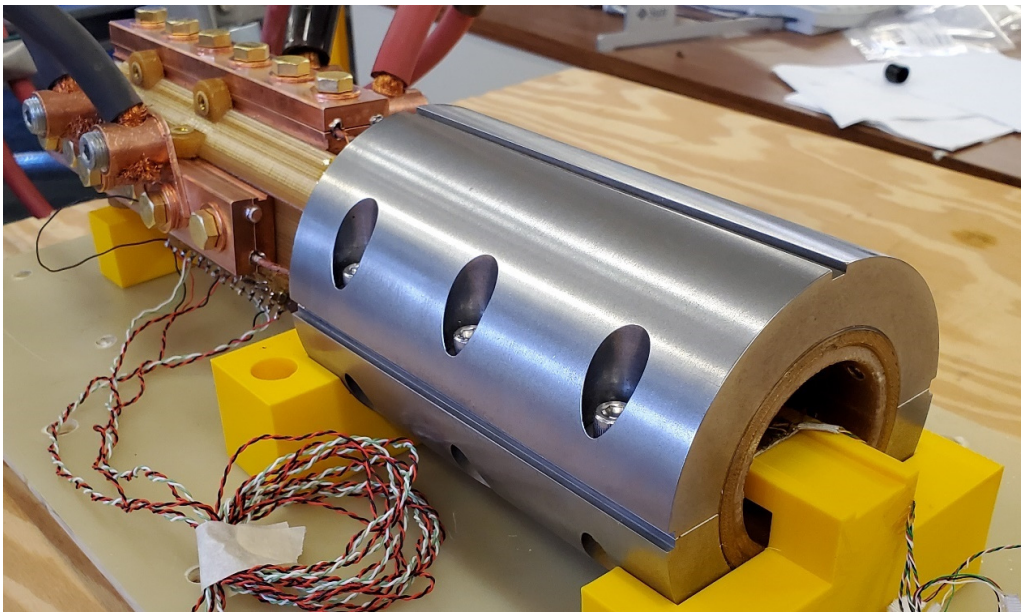

**Figure 8**. Two half-coils assembled with the lead support and the ground insulation (top) and the magnet with the power leads attached prior to testing in liquid nitrogen (bottom).

### 3.3. Magnet testing

The magnet testing consisted of four cool-downs from room to liquid nitrogen temperature, ramping the current to measure the resistive transitions and magnetic field measurements. During the first and second cool-downs, the half-coils were powered in series to produce the intended dipole field shown in figure 6. During the third and fourth cool-down each half coil was powered individually using the middle-point adapter between the coils to gain additional data for a different magnetic field distribution.

The current was ramped with 5-10 A/s ramp rate to the maximum of ~550 A ($I_{max}$), which provided enough data points to measure the resistive transitions. The magnet was not intentionally quenched as it was not the objective of liquid nitrogen testing and since the data acquisition system would not allow to properly detect and characterize the quenches (these studies are planned for the liquid helium tests with a different system), but in several cases the voltage exceeded the safety threshold causing the power supply to shut down. These events did not cause changes in $I_c$ nor n-values measured in subsequent runs.

The typical raw voltage-current characteristics from the magnet tests are shown in figure 9. All voltages are from the voltage taps attached to the outermost layer of the tapes inside of the terminals. The distance between these voltage taps was about 4.4 m. The same resistive criterion of 0.4 µV/cm was applied in all cases after removing the initial wire resistances. Table 1 compares the critical currents achieved during the standalone wire testing with those measured during the magnet and individual coil testing.

STAR® wire 151 performed slightly better in terms of $I_c$ than STAR® wire 131 during the standalone wire testing. However, there was a small but progressive $I_c$ reduction observed in STAR® wire 151 with each current ramp as evident from Table 1. The reason of that behaviour is not fully understood. Since STAR® wire 151 $I_c$ dropped below that of STAR® wire 131 after the coil winding, it suggests that the observed degradation was not due to possible (unmonitored) variation in atmospheric pressure and liquid nitrogen temperature but was rather wire or tape related.

The measured <u>minimum</u> critical current retentions were 99% and 93% for STAR® wires 131 and 151 correspondingly during the magnet testing and there was no $I_c$ reduction with each ramp for either wire. Also, STAR® wire 151 demonstrated a slightly higher $I_c$ retention of 95% during the individual coil testing, which was performed after the whole magnet testing. STAR® wire 131 performed similarly during the magnet and coil testing, demonstrating 99% $I_c$ retention.

The n-value of the wire #151 was higher than that of the wire #131 both during the standalone and the magnet testing. Both wires experienced a small reduction in n-value after the coil winding.

Figure 10 shows the magnetic field distribution on the magnet axis calculated by the 3D model shown in figure 5 and the magnetic field measurements at the corresponding locations. There is a good correlation between calculated and measured data, which means all the turns retained their geometry after the cool-down and energization.

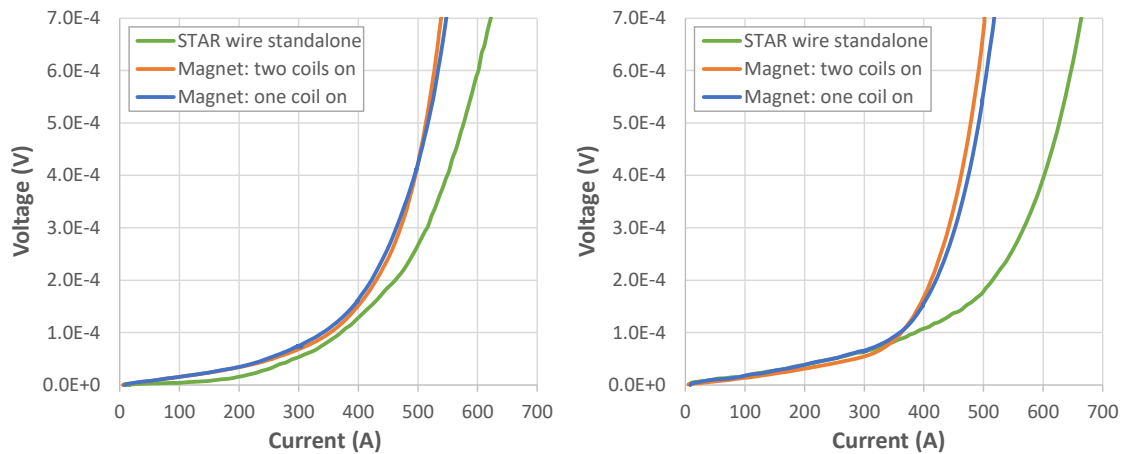

**Figure 9**. Typical voltages of STAR wires 131-151 (left-right) during standalone and magnet testing.

Table 1. Measured and expected wire and magnet performance.

| Test condition | Run # | $I_{max}$ (A) | $I_c$ (A) | n-value | Peak transfer function (T/kA) | Expected $I_c$ (A) | $I_c$ retention (%) |
|---|---|---|---|---|---|---|---|
| STAR® wire 131 standalone | 2 | 681 | 599 | 8.1 | | (595) reference value | (100) reference value |
| | 3 | 776 | 595 | 7.6 | | | |
| | 4 | 819 | 600 | 8.0 | | | |
| | 5 | 851 | 600 | 8.1 | | | |
| | 6 | 834 | 596 | 7.9 | | | |
| | Minimum | 776 | 595 | 7.6 | 0.192 | | |
| STAR® wire 151 standalone | 1 | 806 | 620 | 9.3 | | (606) reference value | (100) reference value |
| | 2 | 836 | 617 | 9.1 | | | |
| | 3 | 836 | 611 | 9.0 | | | |
| | 4 | 801 | 606 | 8.7 | | | |
| | Minimum | 801 | 606 | 8.7 | | | |
| STAR® wire 131 two coils on | 1 | 502 | 455 | 6.4 | | 446 | 99.3 |
| | 2 | 551 | 459 | 7.6 | | | |
| | 3 | 510 | 452 | 6.4 | | | |
| | 4 | 531 | 445 | 6.6 | | | |
| | 5 | 531 | 443 | 6.4 | | | |
| | Minimum | 531 | 443 | 6.4 | 0.694 | | |
| STAR® wire 151 two coils on | 1 | 502 | 425 | 9.5 | | 455 | 92.7 |
| | 2 | 551 | 436 | 9.0 | | | |
| | 3 | 510 | 425 | 7.6 | | | |
| | 4 | 531 | 423 | 7.6 | | | |
| | 5 | 531 | 422 | 7.5 | | | |
| | Minimum | 531 | 422 | 7.5 | | | |
| STAR® wire 131 one coil on | 1 | 550 | 453 | 6.4 | | 457 | 99.1 |
| | 2 | 552 | 454 | 6.3 | | | |
| | Minimum | 550 | 453 | 6.3 | 0.612 | | |
| STAR® wire 151 one coil on | 1 | 532 | 447 | 9.4 | | 466 | 95.2 |
| | 2 | 531 | 444 | 8.0 | | | |
| | Minimum | 531 | 444 | 8.0 | | | |

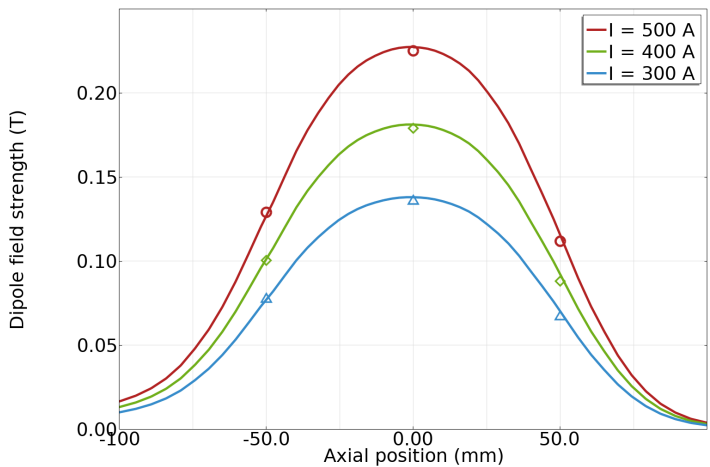

**Figure 10**. Calculated (solid lines) and measured (markers) dipole field on the magnet axis.

## 4. Summary

An HTS dipole magnet with a 60-mm clear bore based on COMB technology was designed, fabricated and tested in liquid nitrogen at Fermilab using STAR® wires produced by AMPeers. The measured critical current retentions for the coils were in 93-99% range, which is an excellent result for any kind of HTS conductor. The magnet went through multiple thermo-cycles and energization cycles without degradation of electrical nor structural properties.

It was the first experimental demonstration of a multi-layer COMB magnet fabricated with ~10 m of REBCO conductor. The results indicate that the COMB magnet technology is compatible with the STAR® wires and allows fabrication of magnets with aperture dimensions relevant for future high energy physics applications. Preparations for testing this magnet in liquid helium are under way.

**Acknowledgments**

This manuscript has been authored by Fermi Research Alliance, LLC under Contract No. DE-AC02-07CH11359 with the U.S. Department of Energy, Office of Science, Office of High Energy Physics, through the US Magnet Development Program.

This work was supported by U.S. Department of Energy Office of Science, Office of High Energy Physics SBIR award DE-SC0022900.